\documentclass[twocolumn]{aastex631}

\graphicspath{ {figures/} }

\newcommand\lmax{\ell_{\text{max}}}
\usepackage[T2A, T1]{fontenc}
\newcommand\sha{\mathop{\text{\normalfont\fontencoding{T2A}\selectfont ш}}}

\usepackage{mathtools}
\usepackage{graphicx, nicefrac}

\usepackage{gensymb} 

\def\mkvec{\overrightarrow{\widehat{M}_k}}
\def\mvec{\overrightarrow{\widehat{M}}}
\def\Sigm1{\Sigma^{-1}}
\def\delqvec{\overrightarrow{\widehat{\delta (q)}}_k}
\def\sumsqinv{\left(G_R^H \Sigm1_R G_R + G_N^H \Sigm1_N G_N \right)^{-1}}

\usepackage[caption=false]{subfig}

\usepackage{amsmath}
\usepackage{amssymb}

\begin{document}
\title{Co-addition and Subtraction of Undersampled Images}

\author{Matan Schlanger}
\affiliation{Department of Particle Physics and Astrophysics\\
Weizmann Institute of Science \\
76100 Rehovot, Israel}
\author{Barak Zackay}
\altaffiliation{Corresponding author: barak.zackay@weizmann.ac.il}
\affiliation{Department of Particle Physics and Astrophysics\\
Weizmann Institute of Science \\
76100 Rehovot, Israel}

\begin{abstract}
In astronomical imaging surveys, repeated observations of the same sky patches are taken in order to obtain deeper images and detect new sources. This is the case in the search for many transient phenomena, such as supernovae, gravitational wave (GW) optical counterparts and other cataclysmic variables. In many such surveys some of the images are undersampled, meaning that the pixel size is too large, and the image suffers from aliasing.
For undersampled images, both co-addition of the images and background subtraction are done in a non-optimal manner, which leads to reduced sensitivity and increased rate of false alarms. \\
We present a new method (named Linear Undersampled Transients \& Addition (LUTRA)) that performs both processes in a mathematically proven optimal way, which allows improved performance for many scientific applications. It also allows easy and direct performance of measurements such as photometry and astrometry in a simple manner, while providing results in super-resolution. We demonstrate the performance of the method on public ZTF data and show $\times 1.25$ higher SNR compared to current methods. We provide an open source Python implementation.

\end{abstract}

\section{Introduction} \label{sec:intro}

\begin{figure}[h]
    \centering
    \includegraphics[width=0.4\textwidth]{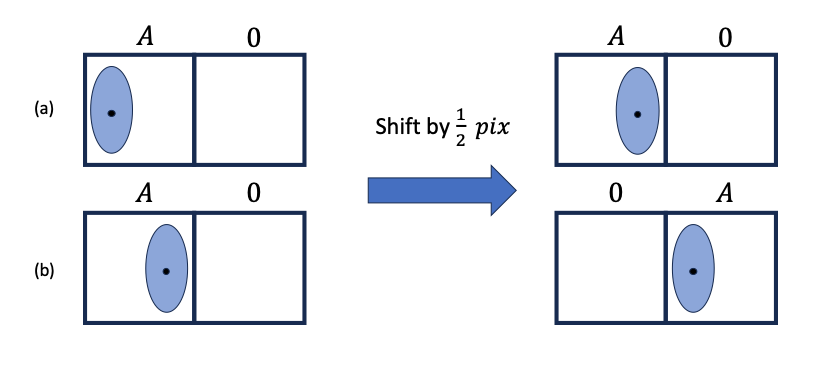}
    \caption{Demonstration of the issues from subpixel shift in undersampled images.\\
            (a) - point source on the left side of a pixel, (b) - point source on the right side of a pixel}
    \label{fig:subpixel_shift_example}
\end{figure}

The field of studying optical transient signals in astronomy, such as supernovae, TDE \citep{Gezari_2021},  gravitational wave (GW) \citep{Bailes2021} counterparts through astronomical imaging is rapidly advancing with more and more efficient ground and space-based imaging surveys being deployed \citep{Ivezic2019-cj, Bellm2018, euclid2025, wfirstfinalreport, McElwain_2023, Ben_Ami_2023, shvartzvald2023ultrasat, chambers2019panstarrs1surveys, Groot_2024}. The primary objectives of these surveys are to obtain deep all sky images for cosmology and to detect and characterize various transient phenomena \citep{lsstsciencebook, euclidsciencebook, spergel2013wfirst24astronomerknow}. \\
During the course of such a survey, researchers typically observe a specific region of the sky and compare it to a set of previous reference images. This comparison aims to identify significant changes that might indicate the occurrence of a transient event. To ensure timely and efficient detection, this process typically involves comparing the new image(s) not to the entire collection of sky images but to a pre-computed summary of them, designed to reduce complexity.\\
For properly sampled images, the process of summarizing reference and subtracting background images was solved in an optimal manner in works such as \citet{zackay2017coaad1}, \citet{zackay2017coaad2} and \citet{zackay2016proper}. However, the methods described in those papers assume that the images are properly registered. 
Achieving registration is straightforward for properly sampled images, where pixel size adheres to the Nyquist-Shannon sampling theorem, which dictates that the pixel size should be smaller than twice the maximum spatial frequency of the signal  \citep{a_course_in_dsp}. For astronomical images, a common rule of thumb is that the Point Spread Function (PSF) of the images should have a width greater than $2$ pixels, denoted as $FWHM>2\ \text{pix}$. Here, FWHM represents Full Width at Half Maximum, which, for a Gaussian signal, approximately equals $2.355 \sigma$.\\
When the Nyquist condition is met, no information is lost in the sampling process \citep{nyquist1928}. 
In the realm of astronomical imagery, the spectral content of an image is generally confined by two primary factors: the optical properties of the telescope, which is prevalent in space-based missions, and the atmospheric turbulence-induced blurring, referred to as seeing conditions, which is the predominant factor in ground-based missions.
A direct correspondence exists between the Fourier transform of a continuous signal and the Discrete Fourier Transform (DFT) of its sampled version. Consequently, a signal can be shifted by any arbitrary amount $\Delta x$ by applying a linear phase shift to the signal's DFT as follows:
\begin{equation*}
    \mathcal{D}\left\{ s(x-\Delta x) \right\} [k] = e^{-2\pi i \Delta x f_k} \mathcal{D}\left\{ s(x) \right\}
\end{equation*}
Here, $\mathcal{D}$ represents the DFT operator and $f_k$ is the frequency corresponding to the index $k$.\\
However, when the condition of proper sampling is not met, the relationship between the continuous signal and the sampled signal becomes more complex, and sub-pixel shifts of the image are not well-defined. Consequently, image registration becomes problematic.\\
A simple case that illustrates this issue is depicted in Figure \ref{fig:subpixel_shift_example}: Suppose we have an image with a PSF that is half a pixel wide, and we encounter two scenarios - one with a source on the left side of the pixel and another with a source on the right side. Shifting the image by half a pixel would have no impact on the first scenario, but it would move the source in the second scenario entirely to the neighboring pixel. However, digital systems are unable to distinguish between the sampled images of both scenarios since they appear identical. Therefore, introducing sub-pixel shifts would require additional, unmeasured information. Sets of images at different sub-pixel shifts may still contain enough information to allow (unobserved) sub-pixel shifts, which provides motivation for this work, aiming to optimally treat sets of undersampled images.\\

\subsection{Current Methods}
One of the most common methods for summarizing undersampled images is the Drizzle algorithm \citep{Drizzle2002}, in which pixels from the various images are mixed into a finer grid by calculating their relative contribution, using a free parameter called the "drop size".  While the Drizzle algorithm generally yields good results, it is not inherently optimal and leaves room for improvement. It also introduces noise correlations that if not accounted for, may result in incorrect inference. Additionally, because it does not leverage information about the PSF, it may encounter challenges when working with images with different PSFs, as it wasn't designed for this scenario. This issue is particularly relevant in many ground-based observations and space-based observations \citep{shvartzvald2023ultrasat} with substantial shifts between images due to differences in PSF and Pixel Response Function (PRF) across different regions of the detector. \\
A different, more rigorous approach was proposed in \citet{Tod_Lauer_1999}. In this approach, the reconstruction of the true sky is performed in the Fourier domain by modeling the images as a linear combination of Fourier components of the sky. By solving the set of equations derived, a high-resolution estimation of the sky depicted in the images is obtained. While this algorithm can be proven optimal in some special cases, it still does not account for differences in PSF between images. Moreover, the algorithm is not numerically stable, as it inverts a matrix that might be singular. Additionally, it weighs the images based on the uniqueness of their relative shift but does not consider the noise level. In other words, shallow images are treated identically to deep images, leaving room for improvement. Last, this approach requires a simultaneous solution for all the images at once, making it challenging to update an existing solution when additional images become available. The entire problem must be solved again. This can also pose a computational challenge for surveys, as there can be as many as $1000$ reference images, and solving for all at once might be a heavy computation. \\
When considering transient detection, it is not initially clear that the problem is well-defined based on a single image, let alone the detection of changes at a resolution smaller than a single pixel. Some approaches, such as \citet{Bramich2012}, involve blurring the images and then applying algorithms designed for properly sampled images. However, this approach has drawbacks, as it sacrifices information due to the blur, requires the specification of another free parameter (the blur radius), and does not address the effects of undersampling.
Other methods, like \citet{Wright2015}, train machine-learning algorithms for transient detection but still rely on human supervision to distinguish real signals from artifacts. \\
Another common approach, used by \citet{Hayden_2021} and \citet{Rubin2021}, involves employing the Drizzle algorithm to co-add the reference images and perform simple arithmetic subtraction between the images (without applying PSF matching). In these special cases, the PSF remains approximately constant between the images. \\
These challenges are not merely technical difficulties. The need for human intervention in transient detection restricts the window of time available for follow-up observations with different instruments and often leads to numerous misdetections and false positives, resulting in the inefficient allocation of valuable observational resources. \\
Another algorithm that is widely in use is the ImCom algorithm \citep{Rowe2011}, famously used in Roman Space Telescope \citep{Hirata_2024}. This algorithm takes a different approach: its objective is a controlled output PSF, achieved by a linear algebra technique. While a constant, known PSF is an important feature for some applications, it directly leads to loss of sensitivity by using sub-optimal co-addition. Moreover, the ImCom algorithm is known to be incredibly demanding in terms of computation and storage \citep{Cao2026}, which limit real-time transient detection.
\subsection{Our work}
In this work, we introduce the Linear Undersampled Transients \& Addition (LUTRA) algorithm—a comprehensive method for synthesizing a collection of undersampled images into a single high-resolution object. This is achieved by calculating a sufficient statistic for any hypothesis regarding the static background. We will demonstrate that this summary retains all the information present in the original data set, ensuring that any hypothesis yields the same likelihood when applied to the summary as it does when applied to the entire data set. The method produces a statistic that facilitates the easy and accurate measurement of astronomical parameters. Furthermore, LUTRA is both numerically stable and updatable, obviating the need to recompute the statistic from scratch when incorporating new data. \\
Moreover, when presented with two sets of images, we provide a statistic for detecting and measuring the amplitude of a new celestial point source. This enables straightforward transient detection. Notably, this statistic is proven to be optimal, solving the problem of transient detection. We will also outline a technique for detecting cosmic rays in images, a crucial component of every astronomical image processing pipeline, and a hard task in undersampled images, as they are very sharp.\\
While we don't treat relative rotation between the images in this paper, this will be treated in a companion paper, introducing a more general approach that allows rotations between the images.
\section{Severity of the problem}
\begin{figure}[h]
    \centering
    \includegraphics[width=0.4\textwidth]{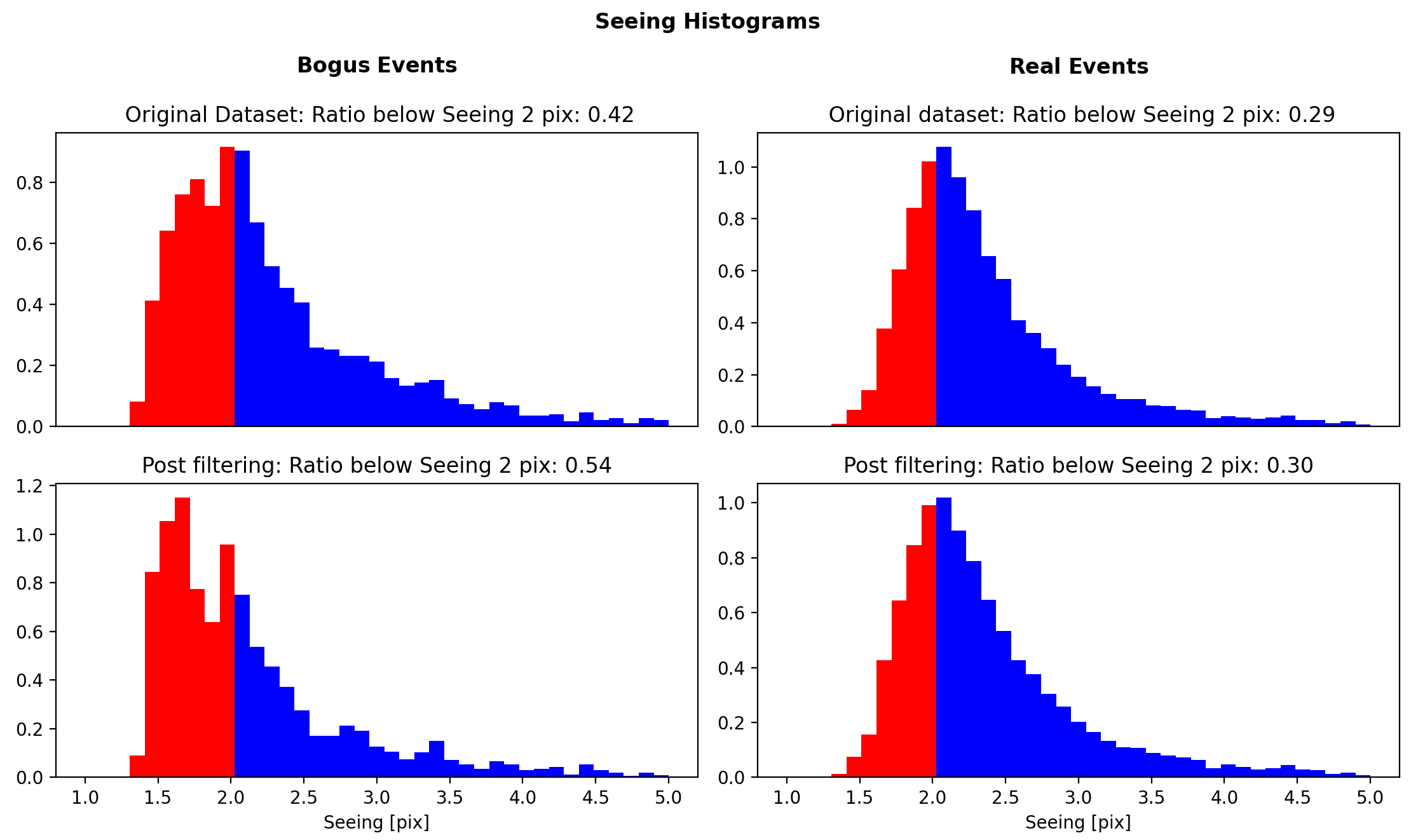}
    \caption{Histogram of the seeing conditions for real and bogus events in \citet{Duev2019}, before and after filtering of events near bright stars.\\
    In red: events from undersampled images.}
    \label{fig:real_bogus_seeing}
\end{figure}

Undersampling is a prevalent issue in various space-based observatories, such as Spitzer \citep{IRAC_2021}, Hubble \citep{hubble1989}, JWST \citep{Jakobsen2022}, Roman \citep{Hirata_2024} and ULTRASAT \citep{shvartzvald2023ultrasat}, as well as ground-based observatories during clear nights, such as BlackGEM \citep{Groot2019-gv}, VRO, where approximately 22\% of images are expected to suffer from undersampling \citep{Ivezic2019-cj}, and ZTF, where roughly 50\% of the images are affected by undersampling \citep{Bellm2018}.\\
In order to evaluate the effects of the current treatment of undersampled images on overall performance, we can analyze the dataset of ZTF events employed in \citet{Duev2019}. This dataset was used to distinguish between genuine transients and spurious events, which are essentially false alarms triggered by the ZTF pipeline. The dataset consists of a total of 11,556 events, with 7,992 classified as real transients and 3,564 categorized as spurious (or bogus) events, all of which were manually classified.
For each class of transient candidates, we construct histograms of the seeing conditions in which these candidates were detected and calculate the relative percentage of events captured in undersampled images. The resulting histogram, displayed in Figure \ref{fig:real_bogus_seeing}, demonstrates that while real events conform to a log-normal distribution (as expected from \citet{Masci2018}), with approximately 30\% of them originating from undersampled images, bogus events exhibit a more pronounced tendency to appear in undersampled images, with around 42\% of them occurring in such conditions.
Furthermore, if we refine our analysis by filtering out events located within less than 20 arcseconds of a star with a magnitude below 16, we observe an even more significant trend. While the histogram for real events barely changes, more than 54\% of bogus events now manifest in undersampled images.
This observation suggests that a substantial proportion of false alarms are attributable to issues arising from undersampling. Therefore, a meticulous approach to addressing these images may lead to an improvement in the false-alarm rate for undersampled surveys.

\section{Modeling}
In this section, we establish a general model for a set of undersampled images, elucidating the relationship between the true celestial sky represented in the image set, the image parameters, and the resultant pixel values. For simplicity, we will employ 1D indices. Whenever the transition to the 2D case is non-trivial, we will explicitly address it.\\
\subsection{Assumptions}
We make the following assumptions:
\begin{itemize}
    \item The PSF of the images is known in advance, with a sufficient resolution better than that of an image pixel. Generally, this can be down by using the images of multiple stars with precisely known positions.
    \item The relative shifts between the images are known to a fraction of a pixel. While better precision yields better performance, it is typically suggested to estimate the relative shift within 0.1 low-resolution pixel.
    \item The noise follows an additive white Gaussian distribution with a known variance. This assumption is valid for background noise and readout noise-limited images, but specialized treatment may be required for areas with bright stars, limited by source noise.
    \item The images are aligned up-to the known shift, with no relative rotation between them. This issue will be dealt with in a future work, where a more general approach will be shown.
\end{itemize}
We will investigate in Section \ref{sec_sensitivity_to_errors} the connection between the precision and the performance.
\subsection{Notation}
\begin{itemize}
    \item We use round parenthesis to indicate a continuous signal and square parenthesis to indicate a discrete signal. For example: $X(\omega)$ is continuous, while $X[n]$ is discrete
    \item We put a hat ($\string^$) to indicate Fourier transform of a signal
    \item We use $k$ to describe low-resolution frequency index and $k'$ to describe high-resolution frequency index
\end{itemize}
\subsection{Sampling}
Let $T(x)$ represent the true sky as depicted in an image set, $P_j$ denote the PSF of the $j$-th image, $\Delta_j$ represent the sub-pixel shift of the image relative to a fixed reference image, and $n_j\sim\mathcal{N}(0,\sigma_j)$ denote the additive white Gaussian noise.

The continuous signal, before sampling, can be expressed as:
\begin{equation}
    m_{j}\left(x\right) = T(x)*P_j(x')*\delta(x'-\Delta_j)
\end{equation}
Here, "*" denotes the convolution operator. \\
Sampling is performed by multiplication with the $\sha$ function, defined as an impulse train \citep{a_course_in_dsp}:
\begin{equation}
M_j(x)=m_j(x)\sha_{f_s}(x), \quad \sha_{f_s}(x)=\sum_{i=-\infty}^{\infty}{\delta(x+i/{f_s})}   
\end{equation}
Here, $f_s$ represents the sampling frequency of the image, which is the inverse of the pixel scale. It's important to note that when discussing frequency, we are referring to \underline{spatial frequency}, not temporal frequency.
In the Fourier domain, the signal becomes:
\begin{equation}
\label{sampled_frequency_relation_with_infs}
\begin{aligned}
    &\widehat{M}(f)=\left(\widehat{T}(f)\widehat{P}_j(f)e^{-2\pi i \Delta_j f}\right)*\sum_{\ell=-\infty}^\infty{\delta\left(f+\ell f_s \right)}=\\
    &\sum_{\ell=-\infty}^\infty{\widehat{T}(f+\ell f_s)\widehat{P}_j(f+\ell f_s)e^{-2\pi i \Delta_j (f+\ell f_s)}}    
\end{aligned}
\end{equation}    
To understand this formula, recall that convolution with a delta function is equivalent to a shift - $g(f)*\delta(f-f_0)=g(f-f_0)$. Thus, the sampled signal consists of an infinite number of replicas of the original signal, each shifted by $\ell f_s$.\\
It is worth noting that while $\widehat{T}(f)$ might have infinite spectral support (as is the case with point sources), the spectral support of the PSFs is inherently finite (i.e., band-limited) by the astronomical seeing, pixel response function (PRF), telescope optics or the finite aperture of the telescope. Therefore, there exists a frequency $f{\text{max}}$ such that for all $j$, $|f|>f_{\text{max}}$ implies $\widehat{P}_j(f)=0$.\\

Recall that when considering the sampled signal, the $k$-th DFT component is the continuous Fourier transform of the sampled signal at frequency $\frac{kf_s}{N}$, where $N$ is the number of points along the transform axis \citep{a_course_in_dsp}: 
\begin{equation} \label{eq_dft_to_regular_fourier}
    \widehat{M_j}[k]=\widehat{M_j}\left(f=\frac{k f_s}{N}\right) \quad\quad k:0...N-1
\end{equation}
Which means that in our analysis, we consider only $0<f<f_s$, as those are the frequencies that will be represented in the sampled signal.
Using this fact, since $\widehat{P}_j(f)$ is band-limited, the sum in Equation \ref{sampled_frequency_relation_with_infs} becomes finite, as if we define:
\begin{equation}
\lmax=\left\lceil\frac{f_{\text{max}}}{\nicefrac{f_s}{2}}\right\rceil-1  
\end{equation}
we can see that 
\begin{equation}
\forall \left|\ell\right|>\lmax, 0<f<f_s: \widehat{P}_j(f+\ell f_s)=0
\end{equation}
We call $B=\lmax+1$ the \textbf{undersampling ratio}.\\
The equation for the sampled signal becomes:
\begin{equation} \label{eq_full_frequency_description}
\begin{split}
    &\widehat{M}_j\left[k\right]= \\
    &\sum_{\ell=0}^{\lmax}\widehat{T}\left(\frac{kf_s}{N}+\ell f_s\right)\widehat{P}_j\left(\frac{kf_s}{N}+\ell f_s\right)e^{-2\pi i \Delta_j \left(\frac{kf_s}{N}+\ell f_s\right)}\\
    &+\widehat{n}_j[k]    
\end{split}   
\end{equation}
For a fixed $k$, $\widehat{M}_j\left[k\right]$ can be regarded as a vector indexed by $j$, denoted as $\mkvec\left[j\right]=\widehat{M}j\left[k\right]$. We also define a vector in $\ell$ as $\overrightarrow{T_k}\left[\ell\right]=\widehat{T}\left(\frac{k f_s}{N}+\ell f_s\right)$ and a matrix $G_{k}\left[j,\ell\right]=\widehat{P}_j\left(\frac{kf_s}{N}+\ell f_s\right)e^{-2\pi i \Delta_j \left(\frac{kf_s}{N}+\ell f_s\right)}$. With these definitions, the samples can be expressed in matrix form for each $k$:
\begin{equation}\label{matrix_form}
\mkvec=G_k\overrightarrow{\widehat{T_k}}+\overrightarrow{\widehat{n}_k}
\end{equation}
The Fourier transform of white Gaussian noise $n\sim\mathcal{N}(0,\sigma)$ is a white complex Gaussian noise, where $\Re\widehat{n}, \Im\widehat{n} \sim \mathcal{N}(0,\sigma\sqrt{\nicefrac{N}{2}})$.\\
This establishes two resolutions for the signals: a low resolution for sampled images and a high resolution where the PSF and the sky are formulated. It's also worth mentioning that a set of vectors in $\ell$ can be transformed into an image by unfolding high-frequency components along the $\ell$ axis, and vice versa by defining a high-resolution index using the transformation:
\begin{equation} \label{eq_high_and_low_res_ind}
    k'=k+N\ell
\end{equation}
This transformation is well-defined, as $k:0,...N-1, \ell:0...J-1, k':0...NJ-1$).\\
From Equation \ref{matrix_form}, we can utilize the fact that $\overrightarrow{\widehat{n}_k}$ is a random complex Gaussian vector (and therefore, so is $\mkvec$). For a given hypothesis of the true sky $T'$, we can directly calculate the log-likelihood as follows:
\begin{equation} \label{log_likelihood_of_hypothesis}
    \begin{aligned}
    &\log{\mathcal{L}\left(T'|\left\{M\right\}\right)}= \log{\Pr\left( \left\{ M\right\} \right)}=\\
    &\sum_{k}\Bigg[-\log\left(\pi^N\det{\Sigma}\right) \Bigg.-\\
    &\Bigg. \left(\mkvec - G_k\overrightarrow{T_k'}\right)^H\Sigma^{-1}\left(\mkvec - G_k\overrightarrow{T_k'}\right)  \Bigg]
    \end{aligned}
\end{equation} 
    
Here, $\Sigma_{i,j}=\mathbf{E}\left[n_i n_j^H\right]=\delta_{i,j}\sigma_j\sqrt{N}$ represents the covariance matrix of the complex white noise in the Fourier domain, and $H$ denotes the Hermitian-conjugate operator.

\section{Statistics Derivation}
In this section, we will utilize the matrix-based approach to derive statistical methods for three key tasks:
\begin{itemize}
    \item Co-addition of an image set, using a sufficient statistic for the likelihood of any hypothesis
    \item Given two image sets, a statistic for the presence of a new point source in an arbitrary position in the sky
    \item Given two image set, a statistic for a pixel being affected by a cosmic ray
\end{itemize}
Our approach will be similar to that employed in references such as \citet{zackay2017coaad1}, \citet{zackay2017coaad2}, and \citet{zackay2016proper}. However, in contrast to treating measurements and the PSF as scalar functions of frequency, we will consider them as frequency-dependent vectors and matrices, respectively.
\subsection{Co-addition} \label{sec_coaddition_derivation}
In this subsection, we explore the decomposition of the likelihood Equation \ref{log_likelihood_of_hypothesis} to derive insights into the statistical methods required for co-addition and detection tasks. We'll apply a Fisher-Neyman factorization approach to identify sufficient statistics for likelihood calculations.\\
We begin by expanding Equation \ref{log_likelihood_of_hypothesis} to isolate two distinct terms:
\begin{equation} \label{eq_log_likelihood_fisher_neyman}
\begin{split} 
    &\log{\mathcal{L}\left(T'|\left\{M\right\}\right)}=\\
    &\underset{h\left(\mkvec\right)}{\underbrace{
        \sum_{k}\left[-\log\left(\pi^N\det{\Sigma}\right) - \mkvec^H\Sigm1\mkvec\right]
        }}+\\
    &\underset{g\left(\mkvec;\overrightarrow{T_k}'\right)}{\underbrace{
        \sum_{k}\left[
            \Re\left\{ \mkvec^H \Sigm1 G_k \overrightarrow{T_k}' \right\}  - \overrightarrow{T_k}'^H G_k^H \Sigm1 G^k \overrightarrow{T_k}'
            \right]
            }}
\end{split}
\end{equation}
This equation reveals that the likelihood can be separated into two parts. The first part does not depend on the hypothesis $T'$ and is, therefore, a constant factor in likelihood calculations. The second part depends on $T'$ only through $\mkvec^H \Sigm1 G_k$ and $G_k^H \Sigm1 G_k$. According to the Fisher-Neyman factorization theorem, these quantities together constitute the "sufficient statistics" for the likelihood of $T'$. This implies that we can compute the likelihood of any hypothesis using only these statistics, without needing the entire data set $\left\{\mkvec\right\}_{j=0}^{J-1}$. \\
Regarding the dimensions of these statistics, $\mkvec^H \Sigm1 G_k$ has dimensions of $H\times W \times B^2$ \footnote{In the 1D case, the dimensions are $N \times B$, as we need separate $\ell$ index per spatial axis, thus giving $B$ per axis}, where $B$ is the downsampling ratio, which can be unfolded using Equation \ref{eq_high_and_low_res_ind} to represent an image with resolution $B$ times finer, having dimensions of $BH \times BW$. The dimensions of $G_k^H \Sigm1 G_k$ are $H \times W \times B^2 \times B^2$, which correspond to the covariance matrix of an image with $B^2 HW$ pixels, as we have. Since typical values for $B$ are 2 or 3, this does not lead to a significant increase in storage requirements relative to the size of a single image.\\ 
To address spatial dependence and efficiently evaluate the likelihood for various positions, consider the hypothesis $T'=d(q)$, representing a source at position $q$. We aim to establish a method to assess the likelihood of such a source at multiple positions efficiently. The likelihood can be expressed as follows:
\begin{equation} \label{eq_log_likelihood_for_delta}
\begin{split}
    &\log{\mathcal{L}\left(T'|\left\{M\right\}\right)}= \\
    &\sum_{k}\Bigg[
            \Re\left\{ \mkvec^H \Sigm1 G_k \widehat{d(q)} \right\}  - \\
            & e^{2\pi i f_{k,\ell} q}\widehat{d(0)}^H G_k^H \Sigm1 G^k \widehat{d(0)}e^{-2\pi i f_{k,\ell} q}
            \Bigg] = \\
    &\sum_{k}\left[
            \Re\left\{ \mkvec^H \Sigm1 G_k \widehat{d(q)} \right\}  - 
            \widehat{d(0)}^H G_k^H \Sigm1 G^k \widehat{d(0)}
            \right]
\end{split}
\end{equation}
Notably, the second term in Equation \ref{eq_log_likelihood_for_delta} does not rely on either the measurements or the position. In many cases, such as source detection, this term can be considered as part of the detection threshold. If we introduce the test statistic as:
\begin{equation} \label{eq_coaddition_test_statistic}
    \overrightarrow{\widehat{S}}_{detection, k} = G_k^H \Sigm1 M_k
\end{equation}
We can observe that:
\begin{equation*}
    \log{\mathcal{L}\left(d(q)|\left\{M\right\}\right)} \sim \sum_k{\Re\left\{ \overrightarrow{\widehat{S}}_{detection, k}^H \widehat{d_k(q)}\right\}} 
\end{equation*}
This allows us to simplify the sum of dot products into a straightforward multiplication in the high-resolution frequency domain, following the procedure outlined in Equation \ref{eq_high_and_low_res_ind}:
\begin{equation*}
    \log{\mathcal{L}\left(d(q)|\left\{M\right\}\right)} \sim \sum_{k'}{\Re\left\{ \widehat{S}_{detection, k'}^* \widehat{d_{k'}(q)}\right\}} 
\end{equation*}
By defining $d'[n]=d[q-n]$ and employing Parseval's theorem, we arrive at:
\begin{equation*}
\begin{aligned}
    &\log{\mathcal{L}\left(d(q)|\left\{M\right\}\right)} = N\Re\left\{\sum_n{ {S}_{detection}[n] d^*(q)[n]}\right\} = \\
    &N\Re\left\{\sum_n{ {S}_{detection}[n] d'(q)[q-n]}\right\} =\\
    &N\Re\left\{{S}_{detection}*d'\right\}[q]
\end{aligned}
\end{equation*}
This establishes that the detection of any source with a template $d[n]$ can be achieved by calculating the matched filter of ${S}_{detection}$ with the template. Hence, we consider ${S}_{detection}$ as a comprehensive summary of the image.
\subsection{Subtraction} \label{sec_subtraction_derivation}
In this subsection, we examine the derivation of statistical methods for image subtraction, which is often used in astronomy to detect transient objects or changes in the sky. The setup involves two sets of images: $M_R$ representing undersampled reference images and $M_N$ representing new images. As the full derivation is extensive, it is provided in Appendix \ref{appndx_full_subtraction_derivation}; here, we present the final result from Equation \ref{eq_s_subtract}. We can define $P_R = G_R^H \Sigm1_R G_R$, $P_N = G_N^H \Sigm1_N G_N$:
\begin{equation} \label{eq_subtraction_stat}
    \begin{split}
        &\widehat{S}_{subtraction} =\\
        &P_R (P_R+P_N)^{-1} G_N^H \Sigm1_N \widehat{M}_N-\\
        & P_N (P_R+P_N)^{-1} G_R^H \Sigm1_R \widehat{M}_R
    \end{split}
\end{equation}

\subsection{Cosmic Ray Detection}
To detect cosmic rays in the new images, consider the setup from Equation \ref{subtraction_setup}. We aim to detect a cosmic ray with amplitude $\alpha$, shape $D$ centered around position $q$. To do this, we need to test the following hypotheses:
\begin{equation} \label{cosmic_ray_hypotheses}
    \begin{split}
        &\mathcal{H}_0: \widehat{M}_N=G_N T_R + n_N \\
        &\mathcal{H}_1: \widehat{M}_N=G_N T_R + \alpha \widehat{D_q} + n_N
    \end{split}
\end{equation}
The derivation is identical to the one in Section \ref{sec_subtraction_derivation}, with $G_N \delta(q)$ substituted with $D(q)$. Similar to Equation \ref{eq_subtraction_beta_log_likelihood}, we get:
\begin{equation} \label{eq_cosmic_ray_beta_log_likelihood}
    \log \beta \sim 2\alpha \Re\left\{ \left(\mvec_N^H-\mvec_R^H C^H\right)\Sigm1_{\mvec_N} \widehat{D}(q)\right\}
\end{equation}
As we've done in Sections \ref{sec_coaddition_derivation} and \ref{sec_subtraction_derivation}, we can define the statistic as:
\begin{equation} \label{eq_cosmic_ray_statistic}
    \widehat{S}_{cosmic-ray} = {\Sigm1_{\mvec_N}}^H \left(\mvec_N-C \mvec_R  \right)
\end{equation}
It's important to note that in this case, a better form of the statistic isn't available. However, since:
\begin{equation*}
    C \mvec_R = G_N\left(G_R^H \Sigm1_R G_R\right)^{-1}G_R^H\Sigm1_R \mvec_R
\end{equation*}
We are still using only the sufficient statistics of the reference images, and we don't need the original reference images.
\subsection{Properties of the Derived Statistics}
\subsubsection*{Optimality}
Both statistics are derived using fundamental statistical theorems like the Neyman-Pearson lemma and the Fisher-Neyman factorization theorem. These theorems ensure the optimality of the statistics, under the assumptions of a known PSF and additive white Gaussian noise with a known variance.
\subsubsection*{High Resolution}
When considering the dimensions of the statistics, it becomes apparent that both are sets of vectors in $\ell$, with one vector per value of $k$. This implies that they can be unfolded into high-resolution images with significantly greater spatial detail. In essence, the statistics provide a resolution improvement by a factor of $B$, where $B$ represents the undersampling ratio.
\subsubsection*{Reproduction of Results}
\begin{figure}[h] 
    \centering
    \includegraphics[width=0.4\textwidth]{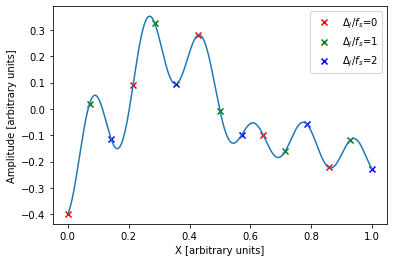}
    \caption{Example of signal interlacing}
    \label{fig:interlacing_example}
\end{figure}

Our statistics replicate previous results that have been proven to be optimal. They are consistent with results in scenarios where straightforward solutions are identifiable. We will explore some of these cases:
\paragraph{Special case - properly sampled images}
In instances where the images are adequately sampled, the matrices and vectors, namely $M, G, T$, simplify to scalars. In such cases, the matrix $G$ represents the PSF, including a linear phase to facilitate image registration. Furthermore, the statistical calculations for co-addition and subtraction reduce to scalar statistics, which have previously been demonstrated as optimal in studies like \citet{zackay2017coaad1} and \citet{zackay2016proper}.
\paragraph{Interlacing}
Consider a scenario where we have $J$ linearly spaced sub-samples of a signal, denoted as $\Delta_j = -\nicefrac{j}{J f_s}$, as depicted in Figure \ref{fig:interlacing_example}. 
Intuitively, we anticipate that the summary of such an image set will resemble the interlaced signal, providing equivalent information to a high-resolution signal. In Appendix \ref{appndx_linearly_spaced_shifts}, we prove that our method indeed achieves this outcome, yielding the same signal up to a constant. This result is akin to the operations of algorithms such as Drizzle \citep{Drizzle2002}, but general to arbitrary number of images and shifts, and without the need for fine-tuning free parameters.

\subsubsection*{Usage of co-addition in subtraction}
Examining Equation \ref{eq_s_subtract}, we observe that the subtraction statistic solely employs terms similar to the co-addition sufficient statistic, including $\widehat{M}_R^H \Sigm1_R G_R$ and analogous terms for the new images. Consequently, by utilizing the co-addition representation of the image sets, we can swiftly compute the subtraction statistic without the necessity of loading the entire data set into memory, including the new images.

\subsubsection*{Numerical Stability}
The co-addition statistic requires only forward calculations, without inverting a matrix that might be singular \footnote{The covariance matrix $\Sigm1$ can't be singular}, and therefore inherently stable numerically. The subtraction statistic requires a careful treatment in order to calculate in a stable manner.The details of our implementation are in Appendix \ref{appndx_stable_subtraction}.
\subsubsection*{Unbiased Estimation}
Upon substituting $\mvec_R=G_R \overrightarrow{\widehat{T_R}} +\overrightarrow{n_R}$ and $\mvec_N=G_N \overrightarrow{\widehat{T_N}} +\overrightarrow{n_N}$ into the subtraction statistic and computing the expected value, we obtain:
\begin{equation}
\begin{split}
    &\mathbb{E}\widehat{S} = \left(\widehat{T}_N - \widehat{T}_R \right) G_N^H \Sigm1_N G_N \\
    &\sumsqinv G_R^H \Sigm1_R G_R
\end{split}
\end{equation}
As evident, the expectation relies directly on the disparity between the true sky representations within the image sets, influenced by an effective PSF. Furthermore, in the scenario where no changes occur in the true sky ($T_R = T_N$), we find that $\mathbb{E}\widehat{S}=0$, suggesting an unbiased estimation of changes.
\subsubsection*{Anti-Symmetry between image sets}
The nature of the image subtraction problem is inherently anti-symmetric. If we were to interchange the new and reference sets, including their corresponding PSFs and noise variances, we would anticipate obtaining the inverse image. This property holds true for our subtraction statistic:
\begin{equation*}
\begin{split}
        & \widehat{S}\left[M_R, G_R, \Sigma_R, M_N, G_N, \Sigma_N \right] = \\
        & - \widehat{S}\left[M_N, G_N, \Sigma_N, M_R, G_R, \Sigma_R \right]
\end{split}
\end{equation*}
as expected from such a statistic. 

\section{Flux and Astrometric Measurements} \label{sec_meas_performance}
In this section, we will discuss how to perform various operations using the statistics we've derived.
\subsection{Background Source Detection} \label{subseq_background_detection}
Consider the case of point-source detection. Recall the likelihood calculation in Equation \ref{eq_log_likelihood_fisher_neyman}. To detect a point source, we aim to distinguish between two hypotheses:
\begin{equation} \label{eq_detection_setting}
\begin{split}
&\mathcal{H}_0: T=0\\
&\mathcal{H}_1: T=A\delta(q)
\end{split}
\end{equation}
According to the Neyman-Pearson Lemma \citep{NeymanPearson1933}, the most powerful statistical test for any given false alarm rate is expressed as:
\begin{equation*}
\log{\Pr\left(M|\mathcal{H}_1\right)}-\log{\Pr\left(M|\mathcal{H}_0\right)}>\eta
\end{equation*}
In this equation, taking the difference helps remove the term in Equation \ref{eq_log_likelihood_fisher_neyman} that is independent of the hypothesis. As demonstrated earlier, we can omit the last term of Equation \ref{eq_log_likelihood_fisher_neyman} from the statistical test and place it in the detection threshold calculation, as it does not depend on the measurements. Consequently, we arrive at the following statistical test:
\begin{equation}
    Y(q)=A\Re\left\{\sum_k{
    \widehat{S}_{detection}^H \delqvec} \right\}
\end{equation}
In this equation, $S_{detection}$ is defined as per Equation \ref{eq_coaddition_test_statistic}, $\widehat{\delta(q)}$ represents the Fourier transform of a single point source at position $q$, and the vector operates in the $\ell$ (replica) dimension. It is important to note that under the assumption of $\mathcal{H}_0$, $M_k$ is Gaussian noise, and $Y$ becomes a Gaussian random variable $R\sim \mathcal{N}\left(0, A\sigma_Y\right)$ with the following standard deviation:
\begin{equation*}
    \sigma_Y = \sqrt{\frac{1}{2}\sum_k{\delqvec^H G^H_k \Sigm1 G_k \delqvec}}
\end{equation*}
The factor $\frac{1}{2}$ arises due to the $\Re$ operation. If we want to determine a threshold $\eta$ such that $\Pr\left\{Y>\eta \right\}<c$ for some $c$, we can employ the approximation:
\begin{equation*}
    \Pr\left(Y>A\sigma_Y n|\mathcal{H}_0\right) \approx \frac{e^{-n^2/2}}{2n\sqrt{\pi/2}}
\end{equation*}
It is important to recognize that both $Y$ and the threshold depend linearly on $A$. Therefore, for simplicity, we can set $A=1$ and test for all $A$ values simultaneously by computing $\sigma_Y$ using the $G_k^H \Sigm1 G_k$ values we already have. We can then select $n$ based on the desired false-alarm rate and test the following condition:
\begin{equation}
    Y(q)=\Re\left\{\sum_k{
    \widehat{S}_{detection}^H \delqvec} \right\}
    >n\sigma_Y
\end{equation}
As previously demonstrated in Section \ref{sec_coaddition_derivation}, we can also efficiently test for all values of $q$ by matched-filtering (or cross-correlating) $S_{detection}$ with the source template and selecting points with significance above the set threshold.

\subsection{Background Photometry}
Revisiting the scenario presented in Equation \ref{eq_detection_setting}, we now aim to estimate the amplitude of the star $A$. By substituting the likelihood of the hypothesis obtained in Equation \ref{eq_log_likelihood_fisher_neyman} and differentiating with respect to $A$ to locate the maximum likelihood, we obtain:
\begin{equation}
\begin{aligned}
    0=\frac{\partial \mathcal{L}}{\partial A} = 
    &\sum_k{
    \Re\left\{\mkvec^H \Sigm1 G_k \delqvec\right\}} - \\
    &A\sum_k{\left\{\delqvec^H G_k^H \Sigm1 G_k \delqvec \right\}} \\
    &\Rightarrow A = \frac{\sum_k{
    \Re\left\{\mkvec^H \Sigm1 G_k \delqvec\right\}}}
    {\sum_k{\left\{\delqvec^H G_k^H \Sigm1 G_k \delqvec \right\}}}
\end{aligned}
\end{equation}
To confirm that this estimation represents the maximum likelihood (rather than the minimum), we can examine the second derivative, and recall that $\Sigm1$ is positive definite:
\begin{equation}
\frac{\partial^2 \mathcal{L}}{\partial A^2}=-\sum_k{\left\{\delqvec^H G_k^H \Sigm1 G_k \delqvec \right\}}<0    
\end{equation}
Recalling the definitions provided in Section \ref{subseq_background_detection} for $Y(q)$ and $\sigma_Y$, we find that the estimated amplitude of a source at position $q$ is given by:
\begin{equation}
    A = \frac{Y(q)}{\sigma_Y^2}
\end{equation}
Therefore, we can readily normalize $Y(q)$ to obtain the estimated photometry at any position. This approach allows for accurate and efficient photometry estimation, benefiting from the statistical properties of the derived co-addition statistic.

\subsection{Precision Astrometry}
To achieve precision astrometry, the following steps are performed:

Let $\Delta$ represent the exact position of the star. The Fourier transform of a point-source located at $\Delta$ is denoted as $T_{k'}=e^{-2\pi i k' \Delta f_s/N}\widehat{\delta (0)_{k'}}$. Substituting this into the likelihood Equation \ref{eq_log_likelihood_fisher_neyman} leads to:
\begin{equation*}
    \mathcal{L}=\sum_{k'}{\Re\left\{
    e^{-2\pi i k' \Delta f_s/N}\widehat{\delta (0)^*_{k'}}\widehat{S_{detection, k'}^*}
    \right\}}=0
\end{equation*}
Using any optimization algorithm, we can minimize this equation to obtain:

\begin{equation*}
    \widetilde{\Delta} \approx \underset{\Delta}{\operatorname{argmin}} \left\{ \mathcal{L} \right\}
\end{equation*}

\subsection{Transient Source Detection}
Assuming no star is present, the statistic derived in Equation \ref{eq_s_subtract} follows a Gaussian distribution. The variance of this statistic can be calculated, as detailed in Appendix \ref{appndx_subtraction_variance}. In summary, for the optimal test:
\begin{equation*}
    \beta = \Re \left\{ \sum_{k'}{\widehat{S}^*_{trans,k'}\widehat{\delta(q)}_{k'}} \right\}
\end{equation*}
As discussed in Section \ref{subseq_background_detection}, this statistic follows a Gaussian distribution with a standard deviation given by:
\begin{equation*}
    \sigma_{\beta'}=\sqrt{\frac{1}{2}\sum_{k'}{
    \left|\widehat{\delta(q)}_{k'}\right|^2   \Re\left\{\Gamma_{\widehat{S},k'}\right\}
    }}
\end{equation*}
The detection threshold can be calculated in a similar manner to what was demonstrated for background detection in Section \ref{subseq_background_detection}.
\section{Estimation of Sensitivity to Parameter Error} \label{sec_sensitivity_to_errors}
In this section, we investigate the sensitivity of our algorithm to errors in various image parameters, including the PSF, noise variance, and image shift. Our goal is to quantify the level of precision required for each of these parameters to ensure the robust performance of our co-addition and transient detection algorithms. \\
For this analysis, we conducted experiments to evaluate the sensitivity of both the co-addition and transient detection aspects of our algorithm to parameter errors. The detailed experimental setup can be found in Appendix \ref{appndx_sensitivity_setup}. These experiments allow us to understand how sensitive our approach is to inaccuracies in parameter estimation and provide insights into the necessary precision for each parameter to maintain reliable performance.
\begin{figure*}[t] 
    \centering
    \includegraphics[width=\textwidth]{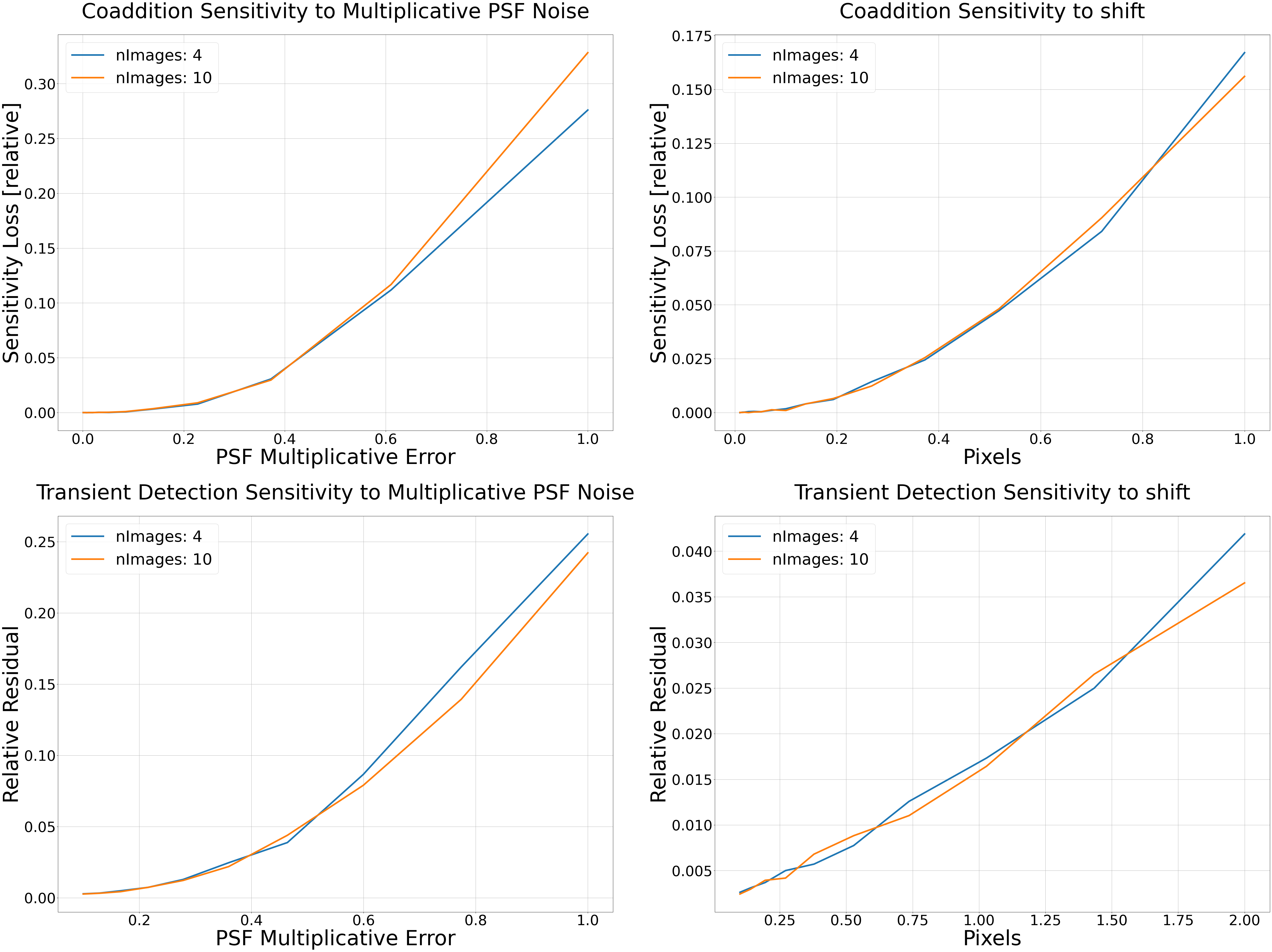}
    \caption{Top row: Transient detection sensitivity (in term of residual, relative to original source) to errors in shift estimation (left) and PSF noise (right)\\
    Bottom row: Co-addition  sensitivity (in terms of loss of sensitivity) to errors in shift estimation (left) and PSF noise (right)}
    \label{fig:parameter_sensitivity}
\end{figure*}
The results, presented in Figure \ref{fig:parameter_sensitivity}, show that for image subtraction, in order to achieve a residual of $<0.01$ from original source, it is required to have the images relative shift to within 0.5 of a pixel and the PSF to within $<0.1$ multiplicative error. It also shows that the loss of sensitivity is independent of the number of images, and can achieve $98\%$ of the sensitivity even within $0.2\ pix$ error

\section{Comparison with Current Algorithms} \label{sec_comparison}
In this section, we will describe how we compare the performance of our statistics with the algorithms that currently in use for the tasks of image co-addition and transient detection using undersampled images. \\
We utilize images from the ZTF \citep{Masci2018}. Since the pixel scale of ZTF is approximately 1", we primarily select science images with a full width at half maximum (FWHM) less than 2". Complete details of the setups are provided in Appendix \ref{appndx_algorithm_comparison_setup}.\\
For image coaddition, we compare the LUTRA co-addition algorithm to Drizzle, a commonly used technique, especially in the context of the Hubble Space Telescope \citep{drizzlepacHandbook}. We implement Drizzle using the Python implementation available from \citet{Spacetelescope}. \\
Regarding transient detection, we compare the LUTRA subtraction algorithm to the approach presented in \citet{Rubin2021}. This approach involves summarizing the reference images using Drizzle and detecting transient signals by performing arithmetic subtraction between the new image and the summary generated by Drizzle.

\subsection{Image Co-addition and Source Detection}
\begin{figure*}[t] 
    \centering
    \includegraphics[width=\textwidth]{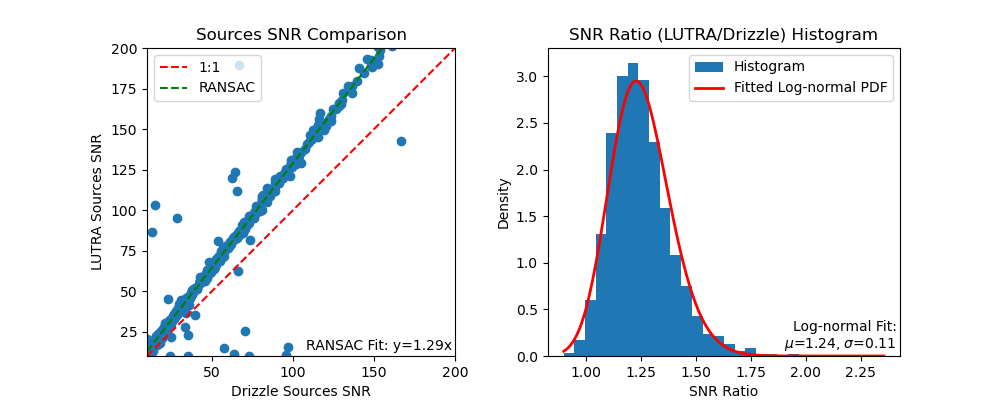}
    \caption{Left: Performance Comparison between LUTRA algorithm and Drizzle for Source Detection for a single experiment (summary of 16 images)\\ Right: Distribution of LUTRA/Drizzle SNR Ratio for multiple experiments}
    \label{fig:lutra_drizzle_coadd_hist}
\end{figure*}

In this section, we conducted a comparison between our LUTRA algorithm and the Drizzle algorithm for image co-addition and source detection. We evaluated the significance of detected sources in both methods and compared their performance.\\
For this comparison, we used images of the arbitrary field centered around coordinates $(RA, Dec) = (314.15926\degree, 27.18282\degree)$. For each iteration, we used 16 images sampled from a collection of 676 images taken by ZTF.\\
We considered the sources detected by the LUTRA statistic and looked for the nearest corresponding sources in the Drizzle results. If there was no matching source within a certain radius, we discarded it. We used the RANSAC algorithm for model fitting in the presence of outliers to account for rare cases where there may be matching errors between the two methods. Full technical details of the comparison are available at Appendix \ref{appndx_coadd_setup}.
\\
The performance comparison, as shown in the left panel of Figure \ref{fig:lutra_drizzle_coadd_hist}, reveals that the significance of sources detected using our LUTRA algorithm is approximately 1.29 times higher than those detected using Drizzle. \\
The specific improvement may vary depending on the choice of images. In order to asses that quantitatively, we ran the experiment 5000 times, each with a different subset of 16 images out of 676, and calculated the distribution of the significance ratio (corresponding to the SNR ratio). In the right panel of Figure \ref{fig:lutra_drizzle_coadd_hist} we can see that the distribution can be fitted as a log-normal distribution, centered around 1.24, thus giving 75\% of samples over 1.15, which corresponds to survey speed improvement of 1.33. \\

While the exact improvement factor depends on multiple parameters, these results demonstrate that our LUTRA algorithm offers a substantial improvement in source detection significance compared to the Drizzle algorithm, making it a valuable tool for image co-addition and source detection in astronomical data analysis.

\subsection{Transient Detection}
In the transient detection experiments, we compared the performance of our LUTRA algorithm with the Drizzle algorithm by injecting of a grid of point sources to a "new" image and measuring its significance. We used images of NGC5154, with 16 reference images and 4 new images, sampled from a total set of 90 images. The images were manually curated in order to remove images with clear stripes in them. Full technical details are available in Appendix \ref{appndx_td_setup}. 
\\

\begin{figure*}[h] 
    \centering
    \includegraphics[width=\textwidth]{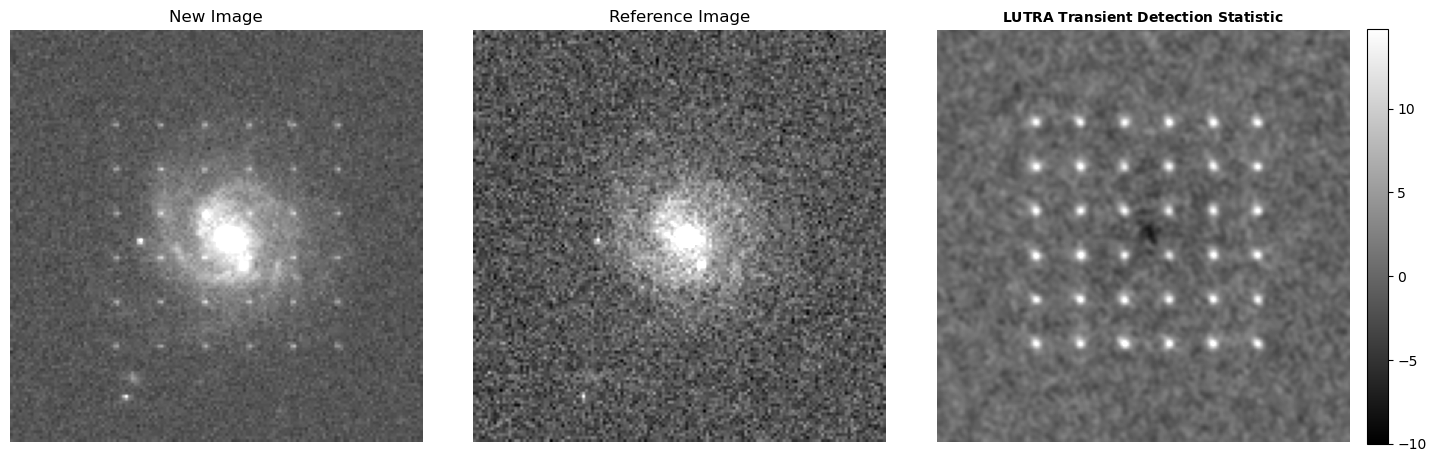}
    \caption{Left to right: Sample new image, sample reference image, LUTRA statistic}
    \label{fig:sn_injection_result}
\end{figure*}

\begin{figure*}[h] 
    \centering
    \includegraphics[width=\textwidth]{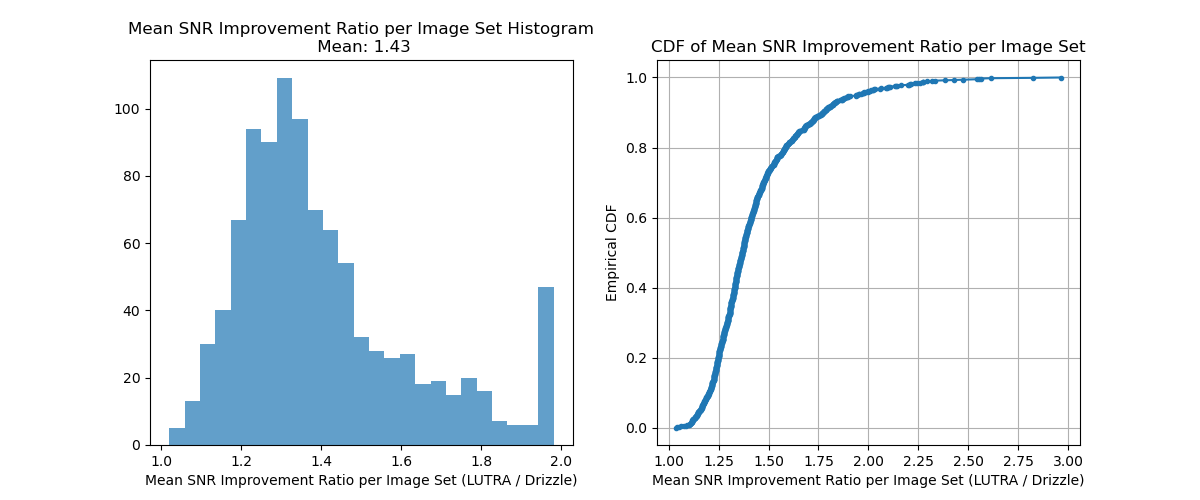}
    \caption{
            Histogram and CDF of Injection SNR Ratio for a faint source}
    \label{fig:sn_injection_hist_and_cdf_combined}
\end{figure*}

In Figure \ref{fig:sn_injection_result}, we can see that our algorithm successfully removes the background galaxy, leaving only the injected sources. Performance is consistent across the different injected sources, without preference to areas closer or further from the galactic center .\\
When comparing the performance of the LUTRA algorithm against Drizzle, we calculate the mean significance of the injected sources after applying each algorithm, and take the ratio between the values for each algorithm. The results in Figure \ref{fig:sn_injection_hist_and_cdf_combined} show  a mean improvement of $\times 1.43$, which corresponds to over +100\% improvement in survey speed for faint transient.

\section{Summary}
The existing algorithms for image co-addition and subtraction suffer from sub-optimal performance, relying on heuristic methods and being susceptible to instability issues that can introduce unwanted artifacts into the processed images. These limitations directly affect the efficiency and cadence of observational surveys. \\
In this paper we have harnessed statistical principles to derive optimal methods for two pivotal tasks in astronomy: image co-addition and transient detection. These tasks serve as the backbone of astronomical investigations, enabling cutting edge research, such as identifying faint transients and precisely tracking the position and brightness of objects within vast survey datasets. \\
Through rigorous testing and comparison, we have demonstrated its effectiveness by showcasing a substantial 1.25-fold improvement in sensitivity compared to the prevailing algorithms used in wide-field surveys. This significant enhancement has the potential to expedite survey speeds by over 1.5 times, a leap forward that can accelerate the detection of transient signals and expedite scientific discoveries for researchers across the globe. \\
One of the standout features of our algorithm is its straightforward, direct computation approach. Unlike iterative methods, our algorithm simplifies the image processing pipeline, reducing the computational demands. This efficiency enables the processing of large images, offering the flexibility to parallelize tasks. This adaptability is particularly crucial for handling the substantial data volumes produced in modern astronomy.\\
We have demonstrated a direct and quantifiable relationship between the precision of input parameters and the performance of the LUTRA algorithm. Specifically, we have shown precisely to what degree each parameter, particularly the relative shift between images and the PSF, needs to be known to achieve sufficient results.\\
While our algorithm has demonstrated optimality in its intended scenario, there remains both theoretical and practical work to be undertaken in order to seamlessly integrate it into future image processing pipelines, including those for missions like the VRO and ULTRASAT. These necessary developments encompass the following key areas:

\begin{itemize}
    \item \textbf{Incorporating Relative Rotation}: Extending the algorithm's capabilities to handle relative rotations between images, a crucial aspect of aligning observations within surveys. 
    
    \item \textbf{Precise Estimation of PSF and Relative Shift at Sub-Pixel Resolution}: Achieving even finer precision in the estimation of the point spread function (PSF) and the relative shift between undersampled images at a resolution finer than a pixel is essential.

    \item \textbf{Pipeline Improvements}: Optimizing the entire data processing pipeline, which includes handling invalid or incomplete data and addressing practical challenges that may arise during large-scale surveys.
\end{itemize}

Addressing these aspects will contribute to the successful implementation of the algorithm in upcoming missions and observatories, allowing astronomers to harness its full potential for image co-addition and transient detection in the pursuit of advancing our understanding of the universe.

\section*{Code Availability}
A basic python implementation, including usage examples and the code that was used to created the figures in the paper, is available  at \url{https://github.com/Zackay-Lab/LUTRA}

\section*{Acknowledgements}
This research was supported by the Israel Science Foundation and the NSF-BSF. B.Z. is supported also by a research grant from the Willner Family Leadership Institute for the Weizmann Institute of Science.
Based on observations obtained with the Samuel Oschin Telescope 48-inch and the 60-inch Telescope at the Palomar Observatory as part of the Zwicky Transient Facility project. ZTF is supported by the National Science Foundation and a collaboration of partners. Operations are conducted by COO, IPAC, and UW
\bibliography{main_bib}{}
\bibliographystyle{aasjournal}

\appendix
\section{Full derivation of Subtraction Statistic} \label{appndx_full_subtraction_derivation}
Both sets are associated with the depicted sky via Equation \ref{matrix_form} for each frequency:
\begin{equation} \label{subtraction_setup}
    \begin{split}
        &\mvec_{R,k} = G_{R,k}\overrightarrow{\widehat{T}}_{R,k}+\overrightarrow{n}_R\\
        &\mvec_{N,k} = G_{N,k}\overrightarrow{\widehat{T}}_{N,k}+\overrightarrow{n}_N
    \end{split}
\end{equation}
Here, $\overrightarrow{n}_R\sim \mathcal{N}(0,\Sigma_R)$ and $\overrightarrow{n}_N\sim \mathcal{N}(0,\Sigma_N)$. \\
The goal is to detect a new point source with amplitude $\alpha$ at position $q$ in the new images. Two hypotheses are tested:
\begin{equation} \label{subtraction_hypotheses}
    \begin{split}
        &\mathcal{H}_0: T_N=T_R\\
        &\mathcal{H}_1: T_N=T_R+\alpha\delta(q)
    \end{split}
\end{equation}
Applying the Neyman-Pearson Lemma \citep{NeymanPearson1933}, the most powerful test is expressed as:
\begin{equation*}
\begin{split}
    &\beta = \frac{\mathcal{L}\left(\mathcal{H}_1|M_R,M_N\right)}{\mathcal{L}\left(\mathcal{H}_0|M_R,M_N\right)} = \frac{\Pr\left(M_N,M_R|\mathcal{H}_1\right)}{\Pr\left(M_N,M_R|\mathcal{H}_0\right)} = \\
    &\frac{\Pr\left(M_N|\mathcal{H}_1, M_R\right)f\left(M_R|\mathcal{H}_1\right)}{\Pr\left(M_N|\mathcal{H}_0, M_R\right)f\left(M_R|\mathcal{H}_0\right)}
\end{split}
\end{equation*}
Since $f\left(M_R|\mathcal{H}_0\right)=f\left(M_R|\mathcal{H}_1\right)$, as the hypotheses don't concern the reference images, we can write:
\begin{equation} \label{eq_beta}
    \beta = \frac{\Pr\left(M_N|\mathcal{H}_1, M_R\right)}{\Pr\left(M_N|\mathcal{H}_0, M_R\right))}
\end{equation}
Using the first equation of \ref{subtraction_setup}, we can express $\widehat{T}_R=\left(G_R^H \Sigm1_R G_R\right)^{-1}G_R^H\Sigm1_R\left(\mvec_R-n_R\right)$ and substitute into the second equation. This leads to:
\begin{equation}
\begin{split}
    &\mvec_N = \underset{C}{\underbrace{G_N\left(G_R^H \Sigm1_R G_R\right)^{-1}G_R^H\Sigm1_R}}\left(\mvec_R-n_R\right)\\
    &+n_N\overset{\text{if } \mathcal{H}_1}{+} \alpha G_N \widehat{\delta(q)}
\end{split}
\end{equation}
$\mvec_N$ is a sum of two Gaussian random vectors and is therefore Gaussian as well. The expected value of $\mvec_N$ and its covariance are given by:
\begin{equation} \label{eq_subt_mn_exp_cov}
    \begin{aligned}
    \mathbb{E}\left[\mvec_N|\mvec_R\right] &= C\mvec_R \overset{\text{if } \mathcal{H}_1}{+} \alpha G_N \widehat{\delta(q)} \\
    \Sigma_{\mvec_N} &= C\Sigma_R C^H + \Sigma_N    
    \end{aligned}
\end{equation}
Calculating the probability densities for Equation \ref{eq_beta} gives:
\begin{multline*}
        \log \Pr\left(\mvec_N|\mathcal{H}_0, \mvec_R\right) = -\left( \mvec_N-C\mvec_R\right)^H\Sigm1_{\mvec_N}\left( \mvec_N-C\mvec_R\right)\\
    \log \Pr\left(\mvec_N|\mathcal{H}_1, \mvec_R\right) = -\left( \mvec_N-C\mvec_R-\alpha G_N \widehat{\overrightarrow{\delta}}(q)\right)^H\Sigm1_{\mvec_N}\left( \mvec_N-C\mvec_R-\alpha G_N \widehat{\overrightarrow{\delta}}(q)\right)
\end{multline*}
By subtracting the equations and removing terms that do not depend on the data, we obtain:
\begin{equation} \label{eq_subtraction_beta_log_likelihood}
    \log \beta \sim 2\alpha \Re\left\{ \left(\mvec_N^H-\mvec_R^H C^H\right)\Sigm1_{\mvec_N}G_N \widehat{\overrightarrow{\delta}}(q)\right\}
\end{equation}
It is worth noting that the test is linear in $\alpha$, allowing us to test for all values of $\alpha$ simultaneously. We define:
\begin{equation} \label{eq_s_subtraction_h_def_with_c}
    \widehat{S}_{subtract}^H=\left(\mvec_N^H-\mvec_R^H C^H\right)\Sigm1_{\mvec_N}G_N
\end{equation}
To establish the direct dependence on parameters such as $G_N, G_R, \Sigma_N, \Sigma_R$, we examine $\Sigm1_{\mvec_N}$. Looking at the first term of the covariance from Equation \ref{eq_subt_mn_exp_cov}, after substituting $C$:
\begin{equation*}
    C\Sigma_R C^H = G_N\left(G_R^H \Sigm1_R G_R\right)^{-1} G_N^H
\end{equation*}
Using the Woodbury Identity \citep{Petersen2008}, we can express $\Sigm1_{\mvec_N}$ as:
\begin{equation*}
    \Sigm1_{\mvec_N} = \Sigm1_N - \Sigm1_N G_N \sumsqinv G_N^H \Sigm1_N
\end{equation*}
We can write the term $\Sigm1_{\mvec_N} G_N$ as:
\begin{multline*}
    \Sigm1_{\mvec_N} G_N = \Sigm1_N G_N- \Sigm1_N G_N \sumsqinv G_N^H \Sigm1_N G_N = \\
    \Sigm1_N G_N \sumsqinv \left[ \left(G_R^H \Sigm1_R G_R + G_N^H \Sigm1_N G_N \right) - G_N^H \Sigm1_N G_N\right] = \\
    \Sigm1_N G_N \sumsqinv G_R^H \Sigm1_R G_R
\end{multline*}
Similarly, when considering the term $C^H \Sigm1_{\mvec_N} G_N$, we can use the matrix identity $A(A+B)^{-1}B=B(A+B)^{-1}A$ to get:
\begin{multline*}
    C^H \Sigm1_{\mvec_N} G_N = \\
    \Sigm1_R G_R \left( G_R^H \Sigm1_R G_R \right)^{-1} G_N^H \Sigm1_N G_N \sumsqinv G_R^H \Sigm1_R G_R =\\
    \Sigm1_R G_R \sumsqinv G_N^H \Sigm1_N G_N
\end{multline*}
Substituting these results into Equation \ref{eq_s_subtraction_h_def_with_c}, we obtain our full statistic:
\begin{equation} \label{eq_s_subtract}
    \begin{split}
        \widehat{S}_{subtraction} = & G_R^H \Sigm1_R G_R \sumsqinv G_N^H \Sigm1_N \widehat{M}_N-\\
        & G_N^H \Sigm1_N G_N \sumsqinv G_R^H \Sigm1_R \widehat{M}_R
    \end{split}
\end{equation}
As demonstrated in Section \ref{sec_coaddition_derivation}, we can detect transients in any position by matched-filtering $\widehat{S}_{subtraction}$ with the corresponding template, typically a point source.
\section{Summary of images with linearly-spaced shifts} \label{appndx_linearly_spaced_shifts}
Consider the case of a set of images $\left\{ M_j\right\}_{j=0}^{J-1}$ with linearly-spaced shifts, denoted as\footnote{The minus sign is due to the fact that a sampling in a shifted position is equivalent to camera shift in the opposite direction} $\Delta_j=-\nicefrac{j}{J f_s}$, we can make several key observations. Here, $J$ corresponds to the downsampling ratio $B$, with $J=B^2$ in the two-dimensional case. We assume that the PSF is an impulse ($\widehat{P}_j = 1$), and the noise has the same variance in every image ($\sigma_j=\sigma$).
To analyze the interlaced signal, we first construct a $J$-expanded signal:
\begin{equation*}
    M_{(\uparrow J)}[k'] = 
    \begin{cases}
        M[k'/J] & k'\%J=0 \\
        0 &\text{otherwise}    
    \end{cases}
\end{equation*}
The interlaced signal, denoted as $\Tilde{M}[k']$, is then obtained by summing over these expanded signals:
\begin{equation*}
    \Tilde{M}[k'] = \sum_{j=0}^{J-1}{M_{j,\uparrow J}[k'-j]}
\end{equation*}
To analyze this signal, we use Discrete Time Fourier Transform (DTFT). The DTFT of an expanded signal is expressed as \citep{a_course_in_dsp}: 
\begin{equation*}
    \widehat{M}_{\uparrow J}\left(\theta\right) = \widehat{M}\left(J\theta\right)
\end{equation*}
Considering the DTFT of the full signal, including the shifts, we have:
\begin{equation*}
    \widehat{\Tilde{M}}(\theta) = \sum_{j=0}^{J-1}{\widehat{M}_j(J\theta)e^{-i\theta j}}
\end{equation*}
To obtain the DFT of the signal, we evaluate the DTFT at the frequency $\theta=\frac{2\pi k'}{N}$:
\begin{equation} \label{eq_interleaved_signal_dft}
    \widehat{\Tilde{M}}[k'] = \sum_{j=0}^{J-1}{\widehat{M}_j\left(\theta=J\frac{2\pi k'}{N}\right)e^{-i2\pi j k'/(JN)}}
\end{equation}
Here, $N$ represents the number of samples in each image, and $JN$ is the total number of samples.
When looking at the statistic's conjugate (which is equivalent), we get:
\begin{equation*}
    R\left[k,\ell\right] = G^H_k\Sigma^{-1}M_k=\frac{1}{\sigma^2}\sum_{j=0}^{J-1}M_j\left(\theta=\frac{2\pi k}{N}\right)e^{-2\pi i \frac{j}{J}\left(\frac{k+\ell N}{N}\right)}
\end{equation*}
To compare the signals, we unfold the extra dimension by using Equation \ref{eq_high_and_low_res_ind}. Importantly, since the DTFT is $2\pi$-periodic, we observe that:
\begin{equation}
    M_j\left(\theta=\frac{2\pi k}{N}\right)=M_j\left(\theta=\frac{2\pi k'}{N}-2\pi\ell \right)= M_j\left(\theta=\frac{2\pi k'}{N}\right)
\end{equation}
This leads us to:
\begin{equation}
    R'[k']=\frac{1}{\sigma^2}\sum_{j=0}^{J-1}M_j\left(\theta=\frac{2\pi k'}{N}\right)e^{-2\pi i jk'/(JN)}
\end{equation}
In conclusion, we obtain the same signal in both cases, up to a constant factor:
\begin{equation}
    \widehat{\Tilde{M}}[k']=\sigma^2 R'[k']
\end{equation}
\section{Calculating the subtraction statistic in a stable manner}
\label{appndx_stable_subtraction}
Recall the expression for $\widehat{S}{subtract}$ from Equation \ref{eq_s_subtract}:
\begin{equation}
    \begin{split}
        \widehat{S}_{subtraction} = &G_R^H \Sigm1_R G_R \sumsqinv G_N^H \Sigm1_N \widehat{M}_N-\\
        & G_N^H \Sigm1_N G_N \sumsqinv G_R^H \Sigm1_R \widehat{M}_R
    \end{split}
\end{equation}
At first glance, the statistic is unstable, as $G_R^H\Sigm1 G_R + G_N^H \Sigm1 G_N$ might be singular, as both $G_R, G_N$ might be almost zero, especially in higher frequencies, where the PSF is near zero. However, the term as a whole can be stable, as every infinite eigenvalue of $\sumsqinv$ will get canceled by $G_R^H\Sigm1 G_H$ or $G_N^H \Sigm1 G_N$. 
To mitigate these effects, we calculated $\sumsqinv$ using the following procedure:
\begin{enumerate}
    \item Calculate the SVD decomposition $USV^H=G_R^H \Sigm1 G_H + G_N^H \Sigm1 G_N$ 
    \item Calculate $S^{-1}$ 
    \item Calculate $S'$ by taking the values of $S^{-1}$, and zeroing every singular value above some threshold (defaults to $10^7$)
    \item Calculate $U^{-1}=U^H$ and $V$
    \item Return $VS'U^H$
\end{enumerate}
Additionally, we allow adding small white noise to the PSF before calculating the $G$ matrices, so at no frequency $G$ will have values near zero. The noise is weaker than the PSF by many order of magnitude (at least 8), so it doesn't change the results of the algorithm. This regularization helped stabilize earlier implementations, even though it's not used in the final results.
\section{Calculating the variance of subtraction statistic} \label{appndx_subtraction_variance}
In this section, we will calculate the variance of the subtraction statistic, denoted as $\widehat{S}{subtract}$, used for transient detection. Recall the expression for $\widehat{S}{subtract}$ from Equation \ref{eq_s_subtract}:
\begin{equation}
\begin{split}
    \widehat{S}_{subtract} = 
    &\underset{S_1}{\underbrace{G_R^H\Sigm1_R G_R \sumsqinv G_N^H \Sigm1_N M_N}} - \\
    &\underset{S_2}{\underbrace{G_N^H\Sigm1_N G_N \sumsqinv G_R^H \Sigm1_R M_R}}
\end{split}
\end{equation}
We also have the statistical test, given by:
\begin{equation*}
    \beta = 2\alpha \Re\left\{ \sum_{k'}{\widehat{S}^*_{subtract,k'}\widehat{\delta(q)}_{k'}}\right\}
\end{equation*}
Now, we calculate the covariance matrix of $S_1$, denoted as $\Gamma_{S_1}$:
\begin{equation*}
    \Gamma_{S_1}=G_R^H\Sigm1_R G_R \sumsqinv G_N^H \Sigm1_N G_N \sumsqinv G_R^H \Sigm1_R G_R
\end{equation*}
And for $S_2$, denoted as $\Gamma_{S_2}$:
\begin{equation*}
    \Gamma_{S_2}=G_N^H\Sigm1_N G_N \sumsqinv G_R^H \Sigm1_R G_R \sumsqinv G_N^H \Sigm1_N G_N
\end{equation*}
Since $S_1, S_2$ are independent, the covariance matrix of $\widehat{S}$, denoted as $\Gamma_{\widehat{S}}$, is given by $\Gamma_{\widehat{S}}=\Gamma_{S_1}+\Gamma_{S_2}$, and using the matrix identity $A(A+B)^{-1}B=B(A+B)^{-1}A$ most terms cancel out and we are left with:
\begin{equation*}
    \Gamma_{\widehat{S},k'} = G_N^H\Sigm1_N G_N \sumsqinv G_R^H \Sigm1_R G_R
\end{equation*}
Now, we calculate the standard deviation of $\beta$, denoted as $\sigma_\beta$:
\begin{equation*}
    \sigma_\beta=2\alpha*\sqrt{\frac{1}{2}\sum_{k'}{
    \left|\widehat{\delta(q)}_{k'}\right|^2   \Re\left\{\Gamma_{\widehat{S},k'}\right\}
    }}
\end{equation*}
We can simplify this expression by removing the $2\alpha$ factor from both the test and the variance. Therefore, for $\beta'=\nicefrac{\beta}{2\alpha}$, the standard deviation is given by:
\begin{equation*}
    \sigma_{\beta'}=\sqrt{\frac{1}{2}\sum_{k'}{
    \left|\widehat{\delta(q)}_{k'}\right|^2   \Re\left\{\Gamma_{\widehat{S},k'}\right\}
    }}
\end{equation*}
This result is not specific to a point source but can be applied to any template used for detection.
\section{Algorithm Comparison Setup} \label{appndx_algorithm_comparison_setup}
\subsection{Co-addition and Source Detection} \label{appndx_coadd_setup}
We followed the following setup:
\begin{enumerate}
    \item We selected an arbitrary starfield for this comparison, preferably with few bright sources to minimize the effects of sensor non-linearities and leakage. The starfield was centered around coordinates $(RA, Dec) = (314.15926\degree, 27.18282\degree)$. An example image of the field is shown in Figure \ref{fig:sec_comparison_example}.
    \item A set of images was randomly selected for the comparison. In the results presented, we used 16 images ($J=16$).
    \item For each image, we estimated the sub-pixel shift using the WCS metadata.
    \item The science PSF provided with the images was interpolated to a resolution $\times 2$ finer using spline interpolation.
    \item We calculated the LUTRA test statistic using the interpolated PSF, following Equation \ref{eq_coaddition_test_statistic}.
    \item A Drizzle co-addition image was created using the selected images and their corresponding WCS.
    \item The PSF of the Drizzle image was estimated by aligning peaks from the Drizzle image and taking the median along each pixel.
    \item The test statistic of the Drizzle image was calculated by cross-correlating the Drizzle image with its PSF.
    \item Both statistics were normalized to the same units by dividing the image by the standard deviation of a certain noise patch (the noise patch was checked using a deeper image with $\times 4$ more images, which showed no significant sources in the patch).
    \item The significance of peaks in the co-added images was compared.
\end{enumerate}
\subsection{Transient Detection} \label{appndx_td_setup}
\begin{figure}[h]
    \centering
    \includegraphics[width=\textwidth]{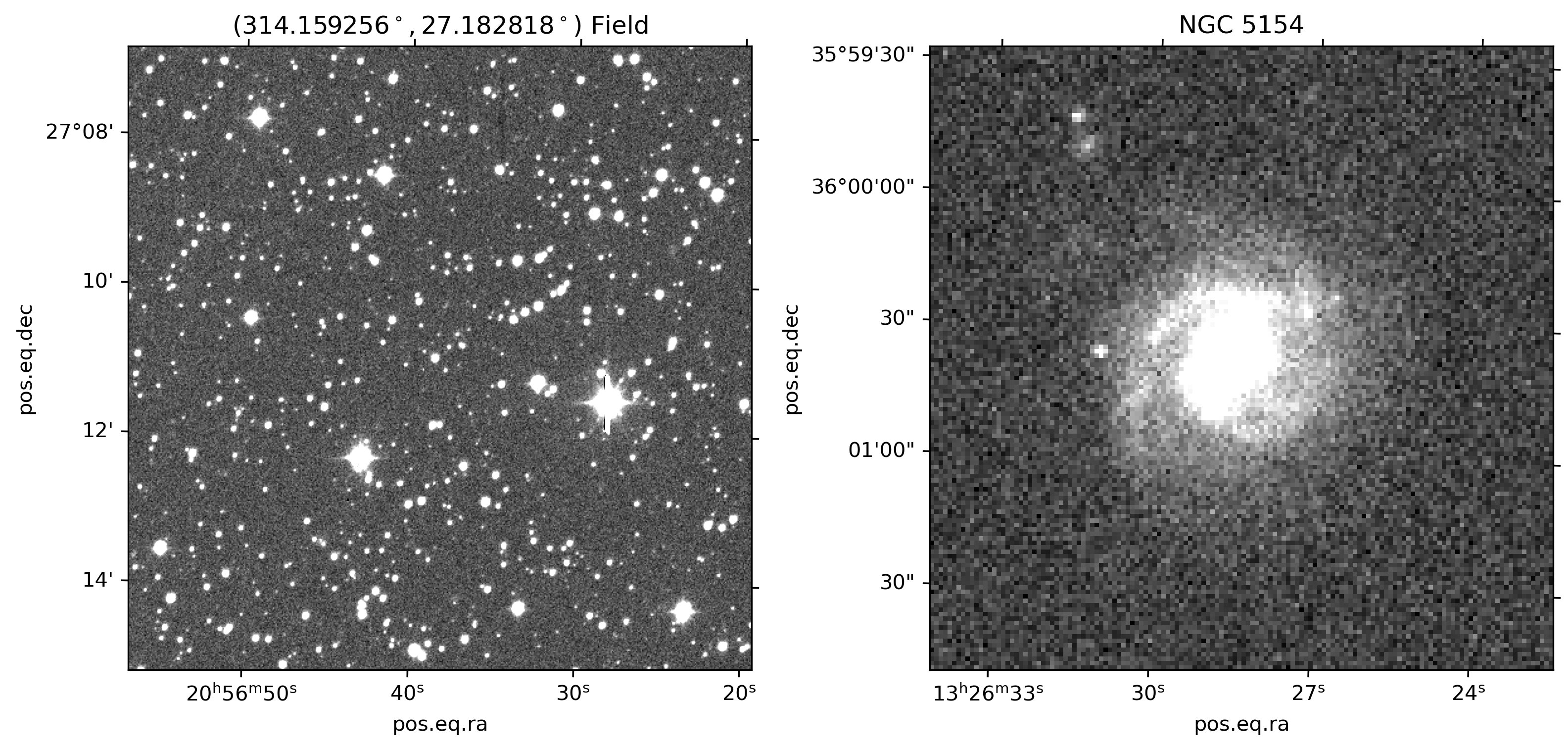}
    \caption{Images of the areas chosen in Section \ref{sec_comparison}}
    \label{fig:sec_comparison_example}
\end{figure}

We followed the following setup:
\begin{enumerate}
    \item We selected an arbitrary galaxy, preferably not too bright to minimize the effects of sensor non-linearities and leakage, and with an active center. Specifically, we chose NGC5154. An example image of the galaxy is shown in Figure \ref{fig:sec_comparison_example}.
    \item A set of images was randomly selected for the comparison.
    \item For each image, we estimated the sub-pixel shift using the WCS metadata.
    \item The images were split into reference images and "new" images.
    \item In the new images, we injected a grid of $6 \times 6$ transient point sources by adding the PSF of the new image, shifted by the relative shift of the image.
    \item LUTRA: 
    \begin{enumerate}
        \item The science PSF provided with the images was interpolated to a resolution $\times 2$ finer using spline interpolation.
        \item The reference images were added to the LUTRA summary.
        \item Transients were detected using the LUTRA test statistic based on Equation \ref{eq_s_subtract}.
    \end{enumerate}
    \item Drizzle:
    \begin{enumerate}
        \item A Drizzle co-addition image for the reference images was created using the images and their corresponding WCS.
        \item A Drizzle co-addition image for the new images was created.
        \item The difference image was calculated by subtracting the reference new Drizzle image from the new image.
        \item Transients were detected in the difference image using the PSF of the new image. In case of multiple images, we used a per-pixel mean between the PSFs.
    \end{enumerate}
    \item Both statistics were normalized to the same units by dividing the image by the standard deviation of a certain noise patch.
    \item The significance of the injected transient was compared over the noise and over the subtraction residuals.
\end{enumerate}
This setup was run with faint injected sources (0.4 times as bright as the brightest pixel), with 16 reference images and 4 new images.

\section{Sensitivity to Parameter Errors} \label{appndx_sensitivity_setup}
\subsubsection*{Co-Addition}
\begin{enumerate}
    \item We created a map of the sky with a fixed number of stars randomly positioned and shaped, all with the same amplitude.
    \item A random PSF width was chosen, and corresponding PSF templates were created.
    \item Images were generated using the true sky and the PSF templates, all with the same noise variance.
    \item The summarization statistic $S_{coadd}$ was calculated using Equation \ref{eq_coaddition_test_statistic} with the exact values for all parameters, normalized to photometry units.
    \item A "distorted" set of parameter values was created by introducing random errors into the parameter values.
    \item The approximated summarization statistic $\Tilde{S}_{coadd}$ was calculated by applying the distorted values to Equation \ref{eq_coaddition_test_statistic}.
    \item The maximal absolute error across the statistic was calculated: $e = \max\left[\left|\Tilde{S}{coadd} - S{coadd}\right|\right]$.
    \item The returned value was normalized by the sources amplitude.
\end{enumerate}
\subsubsection*{Transient Detection}
\begin{enumerate}
    \item Similar to the co-addition scenario, we created a map of the sky with a fixed number of stars randomly positioned and shaped, all with the same amplitude.
    \item A random PSF width was chosen, and corresponding PSF templates were created.
    \item Images were generated using the true sky and the PSF templates, all with the same noise variance. Some images acted as reference images, while one served as a new image.
    \item A "distorted" set of parameter values was created by introducing random errors into the parameter values.
    \item The likelihood of a transient in the image was calculated using Equation \ref{eq_subtraction_stat} with the distorted parameter values.
    \item The maximal residual across the statistic was calculated:
     $e=\max\left[\left|\Tilde{S}_{subtract}\right|\right]$
     \item The returned value was normalized by the sources amplitude.
\end{enumerate}
The default parameters for all the images were as follows:

\begin{itemize}
    \item PSF width: 1.4-1.9 pixels
    \item Stars amplitude: $10^3$ counts
    \item Noise variance: $\sigma=1$ counts
    \item Downsampling ratio: 2
\end{itemize}
Care was taken to ensure that stars did not appear near the image edges to avoid including edge effects in the analysis.

This setup allowed us to assess how sensitive the LUTRA algorithm is to errors in parameter values in both co-addition and transient detection scenarios.

\end{document}